\documentstyle[12pt]{article}
\begin{document}

\setcounter{footnote}{0}
\setcounter{equation}{0}
\setcounter{figure}{0}
\setcounter{table}{0}
\vspace*{5mm}

\newcommand{\alt}{\mathbin{\lower 3pt\hbox
   {$\rlap{\raise 5pt\hbox{$\char'074$}}\mathchar"7218$}}}
\newcommand{\agt}{\mathbin{\lower 3pt\hbox
   {$\rlap{\raise 5pt\hbox{$\char'076$}}\mathchar"7218$}}}

\begin{center}

{\Large\bf Critical indices of Anderson transition:\\
 something is wrong with numerical results} 

\vspace{4mm}
I. M. Suslov \\
P.L.Kapitza Institute for Physical problems \\
117334, Moscow, Russia  \\
E-mail: suslov@kapitza.ras.ru
\\ \vspace{6mm}
\end{center}

\begin{center}
\begin{minipage}{135mm}
{\large\bf Abstract } \\
Numerical results for Anderson transition are critically discussed.
A simple procedure to deal with corrections to scaling is suggested.
With real uncertainties taken into account, the raw data are in 
agreement with a value $\nu=1$ for the critical index 
of the correlation length in three dimensions.

\end{minipage}
\end{center}

\vspace{15mm}

Critical indices $s$ and $\nu$ for the conductivity $\sigma$ and the
localization radius $\xi$ of wave functions near the Anderson transition
\cite{1} are determined by relations
$$
\sigma \sim \tau^s\,\,, \qquad \xi\sim \tau^{-\nu}\,\,,
\eqno(1)
$$
where $\tau$ is the dimensionless distance to a critical point. On the modern 
level, the first estimates of $s$ and $\nu$ were given by a scaling theory of 
localization \cite{2}, $s=1$ and $\nu=1/\epsilon$ for the space 
dimensionality $d=2+\epsilon$. They suggest that $s\approx\nu\approx 
1$  for $d=3$. A little later, a self-consistent theory of Vollhardt and 
W$\ddot o$lfle \cite{3} was developed: it gave for arbitrary 
$d$ \cite{3,4}
$$
s = 1\,\,, \quad d > 2\,\,; 
 \qquad \qquad  \nu = \left \{ \begin{array}{cc}
{\displaystyle \frac{1}{d-2}}\quad , & 2 < d < 4 \\ {   } & {  } \\
{\displaystyle \frac{1}{2}} \quad, & d > 4
\end{array} \right . \quad.
\eqno(2)
$$
These values of indices were in a good agreement with all existing 
information and suspicion arised that they are exact \cite{5}. Later
the present author \cite{6} has suggested a phenomenological scheme,
based on the symmetry considerations, that gives (2) without model
approximations. 

Results based on a nonlinear $\sigma$-model \cite{7} agree with (2)
on a three-loop level, but the correction of order $\epsilon^3$ appears to 
be finite \cite{8} and shifts $s$ and $\nu$ to much lower values, strikingly 
spoiling agreement with experimental and numerical results.  One 
should have in mind, however, that correspondence of a nonlinear 
$\sigma$-model with the initial disordered system {\it is approximate} and 
valid only for $d=2+\epsilon$ with small $\epsilon$. So it is not surprising 
if a difference arises in a certain order in $\epsilon$. On the other hand, 
the high-gradient catastrophe \cite{9} makes the essential modification of 
the $\sigma$-model approach to be inevitable \cite{9a}.

A value $s=1$ is in agreement with a large number of experiments \cite{10}
but this agreement can be doubted on the ground that the interaction effects
are essential in real physical systems. However, attempts to include the 
interaction in the Vollhardt and W$\ddot o$lfle  scheme \cite{11} show that 
a result (2) can preserve in the systems with interaction. Convincing 
evidence for $s=1$ was obtained recently in the experiments with the 
nondegenerate electron gas \cite{12} where the interaction effects are 
negligible.

Early numerical results were in a reasonable agreement with (2) ($\nu=1.2\pm 
0.3$ \cite{13}, $\nu=0.9\pm 0.3$, $\nu=1.4\pm 0.2$ \cite{14}), but later a 
tendency to the larger values arised: $\nu=1.35\pm 0.15$ \cite{15},
$\nu=1.50\pm 0.15$ \cite{15a}, 
$\nu=1.54\pm 0.08$ \cite{16}, $\nu=1.45\pm 0.08$ \cite{17}, $\nu=1.4\pm 0.15$ 
\cite{18}, $\nu=1.58\pm 0.02$ \cite{19}. So large values contradicts to all
other information\,\footnote{\, In a recent communication \cite{20}
A. Kawabata claims that a value $\nu=1.58$ can be obtained, if 
self-consistency of  $q$-dependence of the diffusion coefficient $D$
with $L$-dependence of conductivity is required. In fact, a solution with 
$\nu=1$ is already self-consistent in this sense: it has no $q$-dependence, 
but $L$-dependence of conductivity follows from the temporal dispersion of 
$D$ \cite{3}. The general analysis shows \cite{6} that only a solution with 
negligible $q$-dependence is self-consistent.
} on the critical indices.
\vspace{3mm}

It means, in our opinion, that something is wrong with numerical results.

\vspace{3mm}

1. To understand it, let us consider  $d$-dependence of the numerical
value for $\nu$, which is given by an empirical formula \cite{21}
$$
\nu \approx \,\frac{0.8}{d-2} +0.5
\eqno(3)
$$
This formula summarizes results for $d=3$, $d=4$ and several noninteger 
dimensionalities which are realized in the fractal structures. From the 
theoretical viewpoint, the formula (3) is entirely unsatisfactory:

(a) It gives $\nu=0.8/\epsilon$ for $d=2+\epsilon$, while it should be   
$\nu=1/\epsilon$  in any variant of one-parameter scaling: it is a
consequence of the fact that the Gell-Mann -- Low function has a behavior
$\beta(g)=(d-2)+A/g$ for large $g$, the latter property being proved by a 
diagrammatic analysis. Of course, one can doubt in validity of 
one-parameter scaling, but in such case the whole procedure of 
the data treatment should be dismissed as entirely based on it.

(b) Formula (3) takes into account that $\nu=1/2$ for high dimensions
\cite{22} but suggests the infinite value for the upper critical 
dimensionality $d_{c2}$. Such hypothesis arised in the formalism
of $\sigma$-models, which do not exhibit  any special dimension 
except  for $d=2$ \cite{22a}. Once again, it is a consequence
of the fact that the $\sigma$-model approach is approximate
and can be justified only for low dimensions.\,\footnote{\,Efetov's 
derivation \cite{22a} of the $\sigma$-model is formally valid for
arbitrary dimensions but includes the artificial construction of 
weakly-coupled metallic granules. Such construction
and related with it approximations can have unpredictable influence
on results.}

In the exact field theory formulation, the theory of disordered 
systems is equivalent to the $\varphi^4$ theory with a "wrong" sign of 
interacton \cite{23,23a}. The latter is renormalizable for $d\le 4$ and 
nonrenormalizable for $d>4$ \cite{23b}\,\footnote{\, Problems of the "wrong"
sign of interaction and the replica limit $n\to 0$ are inessential in the 
framework of perturbation theory. In fact, renormalizability can be 
investigated directly in the framework of the "impurity" 
diagrammatic approach
\cite{23c}, without transition to the effective field theory.}. 
For $d\le 4$ all of the physics is 
determined by small momenta or large distances, in accordance with the 
expected scale invariance. For $d>4$ the atomic 
scale cannot be excluded from results and no scale invariance is possible. So 
the upper critical dimension is surely four and it can be seen from the 
different viewpoints \cite{24}.  So we should expect $\nu=1/2$ for $d=4$, but 
it is in conflict with numerical value $\nu=1.1\pm0.2$ \cite{25}, which can 
be doubted on these grounds.  By continuity, the analogous systematic error 
can be expected for $d=3$.

\vspace{3mm}

2. The most probable reason for a systematic error is the existence of
corrections to scaling. All numerical results are based on one-parameter 
scaling and indeed there are no serious doubt in it for low dimensions. 
However, there are no indications of the existence of the upper critical 
dimensionality in the framework of one-parameter scaling. The only 
possibility to resolve this controversy is to suggest that one of irrelevant 
parameters becomes relevant at $d=4$. The corresponding scenario was 
developed in \cite{26} and resulted in the scale dependence of the 
conductance which is in agreement with the Vollhardt and W$\ddot o$lfle  
scheme.  There is two-parameter scaling for $d\ge 4$ and a large systematic 
error for $d=4$ is rather natural. For $d=4-\epsilon$, two-parameter scaling 
takes place for intermediate scales and the region of its applicability 
becomes smaller with increase of $\epsilon$. As a consequence, disagreement 
of numerical results with (2) is less for lower dimensions.

On the model level, corrections to scaling were analysed in \cite{27}. With 
some assumptions, it appears to be possible to agree the raw 
data with a value $\nu=1$.

As it was claimed in \cite{19}, with corrections to scaling taken into 
account, the accuracy of results for $\nu$  reaches the level $\sim 1\%$.
This statement is extremely doubtful. Authors of \cite{19} carry out
nonlinear fitting with 10--12 parameters and present only results but not the
procedure used. But in such situation, $\chi^2$ has a great number of 
different minima (as in a spin glass), and the true one is not necessary the 
deepest:  there are no real possibility to systematize them. Probably, 
authors of \cite{19} have in mind the minimum, nearest to the fit without 
scaling corrections. But a problem is, that such fit may be not a good zero 
approximation. 

\vspace{3mm}

3. We can suggest a simple procedure to deal with corrections to scaling.
As a starting point, we suppose existence of the abstract renormalization 
group, which is determined by the operator $R_s$: it corresponds to a 
change of the length scale by a factor $s$ and transfers one point ${ \mu}$
of the parameter space into another point ${ \mu'}$, $\mu'=R_s \mu$ 
\cite{23}. If $\mu^*$ is a fixed point, $\mu^*=R_s \mu^*$, then we have for 
small $\delta\mu=\mu-\mu^*$
$$
\delta\mu=\mu-\mu^*=\sum\limits_i A_i(\tau) \hat e_i\qquad {\rm and}\qquad
R_s\delta\mu=\sum\limits_i A_i(\tau) s^{y_i}\hat e_i\,\,,
\eqno(4)
$$
where $\hat e_i$ and $s^{y_i}$ are eigenvectors and eigenvalues of the 
operator $R_s$, considered as linear in the vicinity of $\mu^*$ \cite{23}.
If $\mu$ is represented by a set of the coefficients $A_i(\tau)$, then we can 
write for any quantity $Q$ with a zero scaling dimensionality:
$$
Q(\mu)=F\left\{A_1(\tau) s^{y_1},\, A_2(\tau) s^{y_2},\,
     A_3(\tau) s^{y_3},\,\ldots  \right\} \,\,,
\eqno(5)
$$
where $y_1>y_2>y_3>\ldots$ and $y_1$ is related with the critical index $\nu$
of the correlation length, $y_1=1/\nu$ \cite{23}. If  we expand $A_i(\tau)$  
near the critical point, $A_i(\tau)=B_i\tau$, we have after redefinition of 
$F$
$$
Q(\tau,L)=F\left\{\tau (L/a)^{y_1},\, \tau (L/a)^{y_2},\,
     \tau (L/a)^{y_3},\,\ldots  \right\}  \,\,,
\eqno(6)
$$
where a length scale $L$ is $s$ times greater than an atomic scale $a$.
For $\tau (L/a)^{y_1} \ll 1$ all arguments of $F$ are small and
$$
Q(\tau, L)\approx F\left\{0, 0,0,\ldots  \right\} +
 C_1\tau (L/a)^{y_1}+ C_2 \tau (L/a)^{y_2}+ C_3\tau (L/a)^{y_3}+ \ldots
$$
$$
\equiv F\left\{0, 0,0,\ldots  \right\} + \tau f\left(L/a \right)\,\,,
\qquad  \tau (L/a)^{y_1} \ll 1 \,\,.
\eqno(7)
$$
In the case of one relevant parameter we have  $y_1>0$, $y_2,\,y_3,\,\ldots<0$
and for large $L/a$ only the first argument in (6)  produces essential 
changes in $F$, while influence of other parameters are uniformly bounded
by a quantity of order $\tau$:\,\,\footnote{\, We assume that a function $F$ 
and its derivatives are bounded.}
$$
Q(\tau,L)\approx F\left\{\tau (L/a)^{y_1},\,0,\,0,\,\ldots  \right\}
\equiv G\left\{\tau (L/a)^{y_1} \right\} \,\,,
\qquad  \tau (L/a)^{y_1} \agt 1 \,\,.
\eqno(8)
$$
Now we can unify (7) and (8) into one expression:
$$
Q(\tau,L)= G\left\{\tau f(L/a) \right\} \,\,.
\eqno(9)
$$
A function $f$ has a power-law asymptotics for large $L$
$$
f(L/a)\sim (L/a)^{1/\nu}\,\,, \qquad L/a\to\infty
\eqno(10)
$$
and an arbitrary behavior for $L/a\sim 1$, when all terms 
$C_i \tau (L/a)^{y_i}$ in (7) are of the same order. Nevertheless, 
dependence of $\ln f(x)$ on $\ln x$ is slow for a wide class of smooth 
functions and in a restricted interval it can be linearised. So
$$
f(L/a)\approx const \,(L/a)^{1/\nu_{eff}}\,\,
\eqno(11)
$$
and we have a relation
$$
Q(\tau,L)= G\left\{\tau (L/a)^{1/\nu_{eff}} \right\} =
           \tilde G\left\{\xi(\tau)/L \right\}\,\,,
\qquad \xi(\tau)\sim a\tau^{-\nu_{eff}} \,\,,
\eqno(12)
$$
which is suggested in all papers involved in the numerical business.
At $L\to\infty$ index $\nu_{eff}$ tends to $\nu$, but at finite $L$
it may be essentially different.

Expression (9) is more general than (12) (only $\tau\ll 1$ was suggested)
and a simple procedure of the data treatment  can be based on it.
The most extensive information on the function $G$ can be obtained
for a maximum value of $L$, which is available for us. If we accept
$f(L_{max}/a)=1$, fixing the scale of the argument of $G$, we have

$$
G\left\{\tau  \right\}= Q(\tau,L_{max})  \,\,.
\eqno(13)
$$
If $Q(\tau, L)$ is known for a set of the scales $L_i$, the formula (9) 
determines $f(x)$ in the points $L_i/a$: we should fit the scale
$\tau_i$ to a relation $G\{\tau/\tau_i\}\approx Q(\tau,L_i)$ and put 
$f(L_i/a)=1/\tau_i$ after that. Such procedure makes it possible to extract 
the true function $f(L/a)$,  not assuming a power-law dependence for it.
Its behavior for $L\sim L_{max}$ gives the most reliable estimate of $\nu$,
that can be obtained from the existing data.
\vspace{3mm}

4. As an illustration, let us discuss the interpretation of the 
largest scale data, existing for the Anderson transition. The largest
system size $L=100$ was used in \cite{18} (we accept $a=1$ in what follows),
but the suitable data are available only for $L\le 28$. The most detailed of 
them are presented in Fig.~4 of \cite{18}, where  a scaling 
quantity $A$ is plotted against the amplitude of disorder $W$  for 
$L=28,\,12,\,6$; we consider $A$ as $Q(\tau,L)$ with $\tau=(W-W_c)/W_c$, 
where $W_c$ is a critical disorder.  

Theoretically, the simplest procedure (in fact, it was used in \cite{18}) can 
be based on the formula (7), which determined $f(L/a)$ as a slope of a linear 
dependence $Q$ on $\tau$ at $\tau\to 0$. In practice, for sufficiantly small
$\tau$ one should take a difference of two close quantities with a great loss 
of accuracy, while for a moderately small $\tau$ it is difficult to control 
the validity of (7). As one can see in Fig.~4 of \cite{18}, dependence $Q$ on 
$\tau$ is only approximately close to linear, but in fact it 
is an essentially broken line. To demonstrate a situation, we give in a 
Table an average slope of dependence $Q$ on $\tau$ and its fluctuations in 
the interval $16<W<17$, which corresponds to a condition 
$\tau(L/a)^{1/\nu}\alt 1$.

  \begin{center}
\hspace{0mm} {\em T a b l e} \\
 Slope of dependence $Q$ on $\tau$ (arbitrary units)
\vspace{3mm}
\begin{tabular}{||c|c|c|c||} 
\hline
      &  $L=28$    & $L=12$ & $L=6$    \\      
    \hline
Average value   & $0.30$  &  $0.16$  &  $0.10$  \\
    \hline
Least value for $16<W<17$ & $0.20$  & $0.10$  & $0.04$ \\ 
\hline 
Largest value for $16<W<17$ & $0.42$  & $0.25$  & $0.12$ \\ 
 \hline 
\end{tabular} 
\end{center}

With the use of the average slope, we indeed obtain $\nu\approx 1.4$, as it 
was reported in \cite{18}. With real uncertainties taken into account, we can 
have any value of $\nu$ in the interval $0.7\div 3.0$. Authors of \cite{18}
give essentially smaller error, relying on the averaging procedure. But one 
can see by eye in Fig~4 of \cite{18}, that any value, given in Table~1, has 
a reasonable probability to be a true slope at $\tau=0$. Statistical 
treatment cannot improve this situation, but can give only an illusion of 
doing it.

More stable results can be obtained with the use of a formula (9). If we 
accept $f(28)=1$ and take $Q(\tau,28)$ as $G(\tau)$, we can find a scaling 
factor $\tau_i$ for each experimental point for $L=12$ and $L=6$. For a 
function $f(x)$ we have
$$
f(12)=0.33\div 0.65\,,\qquad f(6)=0.14\div 0.40
\eqno(14)
$$
and after averaging
$$
\langle f(12)\rangle=0.52\,,\qquad \langle f(6)\rangle=0.28\,\,.
\eqno(15)
$$
One can see in Fig.~1, that the most probable value is $\nu=1.25$,
and it is essentially shifted\,\footnote{\,The origin of this systematic 
shift is rather simple. The function $Q(\tau)$ is linear for small $\tau$
but grows more slowly, something like $|\tau|^\alpha {\rm sign} \tau$
with $\alpha<1$, for larger $\tau$. If the latter dependence is approximated
by linear one, using values for $\tau=\tau_0$ and $\tau=-\tau_0$, the slope
$\sim\tau_0^\alpha$  is obtained. After a scale transformation $\tau\to
\tau (L/a)^{1/\nu}$ this slope has dependence $(L/a)^{\alpha/\nu}$, which is
interpreted as $(L/a)^{1/\nu_{eff}}$ with $\nu_{eff}=\nu/\alpha > \nu$.
Such effect is present in the most of papers.  } in comparison with $\nu=1.4$ 
given in \cite{18}.  With the total uncertainty taken into account, we have 
$\nu=0.8\div 1.7$ and a value $\nu=1$ is admissible.

Uncertainties of results can be formally diminished, if we interpret an error 
in the mean square deviation or another sense. However, it requires some 
statistical hypothesis concerning a distribution of errors, which hardly
can be justified in view of their partially systematic nature. 
\vspace{3mm}

5. An effective index $\nu_{eff}$ in (11,12) tends to $\nu$ in the limit
$L\to\infty$. For the moderate $L$, a good accuracy of approximations (11,12)
does not mean, in general, that $\nu_{eff}$ is really close to $\nu$.
A well known example is the second order transition with a small Ginzburg
number, where a new scale $\xi_0\gg a$ arises: the true critical indices are 
observed only for $L\agt\xi_0$, while for $L\alt\xi_0$ the relations 
(11,12) are valid with a mean-field value of $\nu_{eff}$. In fact, 
convergence $\nu_{eff}$ to $\nu$ can be slow even in 
the absence of such new scale. Evidently, there are no rigorous procedure to 
control a difference $\nu_{eff}-\nu$.

All  numerical estimates of $\nu$ are systematically greater than unity and, 
beyond any doubt, it reflects a realistic behavior of $f(L/a)$
in the corresponding range of scales. A function $f(L/a)$ is not universal
and depends on the choice of a model. One should have in mind, however, 
that one and the same model is investigated in all papers: it is the Anderson 
model with a transition in the center of the band. Differences in the form of 
distribution of cite energies (f.e. box or Gaussian) are of minor importance:
a critical value of $W$ is large, and near the band center this distribution 
looks as practically uniform.  To have a real estimate of a systematic error,
one should shift to a band edge and use the different procedures to reach a
critical point (f.e. by change of energy or disorder). The only attepmpt of 
this kind was undertaken in \cite{14} and resulted in essential decrease
of $\nu$ with increase of uncertainties. It is desirable to repeat such
attempt on the higher level of accuracy.
\vspace{3mm}

In conclusion, we have demonstrated, that even the most detailed and
largest scale data can be reinterpreted with an essential shift of 
results. As for the systems of smaller size, the results can be doubted 
simply on the ground, that the function $f(L/a)$ does not reach its 
asymptotic regime (10). 
As for a systematic error, nobody can 
say nothing definite about it\,\footnote{ \,The most extremistic values of $\nu$
\cite{16,19} are obtained in the case of small $L$ and high accuracy. 
It is exactly the situation, when a statisical error is small, while
a systematic one is large. Surely, the given accuracy reflects only the first 
of them.}. 
\vspace{3mm}

The author is greatful to M. V. Sadovskii for critical remarks.
\vspace{3mm}

This work was financially supported by  INTAS (Grant 99-1070) and the 
Russian Foundation for Basic Research (Project 00-02-17129).

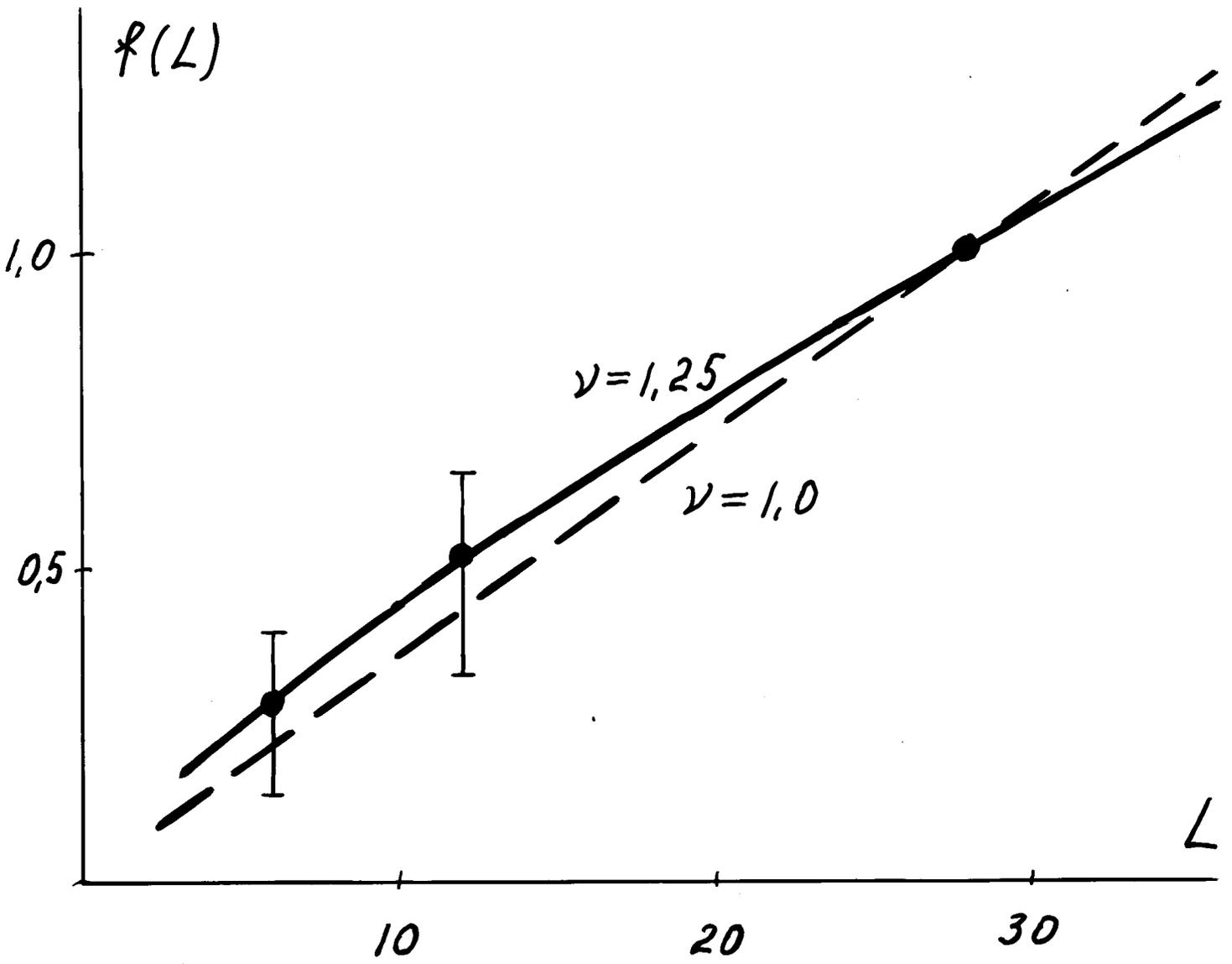
\end{document}